\documentclass{article}

\usepackage{arxiv}

\usepackage[utf8]{inputenc} 
\usepackage[T1]{fontenc}    
\usepackage{hyperref}       
\usepackage{url}            
\usepackage{booktabs}       
\usepackage{amsfonts}       
\usepackage{nicefrac}       
\usepackage{microtype}      
\usepackage{graphicx}
\usepackage{doi}

\usepackage{adjustbox} 
\usepackage{multirow}
\usepackage{graphicx}
\usepackage{amsmath}
\usepackage{amstext}
\usepackage{amssymb}
\usepackage{subcaption} 

\title{A Data-Centric Approach for Safe and Secure Large Language Models against Threatening and Toxic Content}

\date{April 14, 2025} 					

\author{
        Chaima Njeh \\
	  Quebec University at Chicoutimi \\
	  Quebec, Canada \\
	  \texttt{chaima.njeh@yahoo.com} \\ 
	  \AND
	  Haïfa Nakouri \\
	  Quebec University at Chicoutimi \\
	  Quebec, Canada \\
	  \texttt{hnakouri@uqac.ca} \\
	\And
	Fehmi Jaafar \\
	Quebec University at Chicoutimi \\
	  Quebec, Canada \\
	\texttt{fjaafar@uqac.ca} \\
}

\renewcommand{\shorttitle}{\textit{arXiv} Template}

\hypersetup{
pdftitle={A template for the arxiv style},
pdfsubject={q-bio.NC, q-bio.QM},
pdfauthor={David S.~Hippocampus, Elias D.~Striatum},
pdfkeywords={First keyword, Second keyword, More},
}

\begin{document}
\maketitle

\begin{abstract}
Large Language Models (LLM) have made remarkable progress, but concerns about potential biases and harmful content persist. To address these apprehensions, we introduce a practical solution for ensuring LLM's safe and ethical use. Our novel approach focuses on a post-generation correction mechanism, the BART-Corrective Model, which adjusts generated content to ensure safety and security. Unlike relying solely on model fine-tuning or prompt engineering, our method provides a robust data-centric alternative for mitigating harmful content.
We demonstrate the effectiveness of our approach through experiments on multiple toxic datasets, which show a significant reduction in mean toxicity and jail-breaking scores after integration. Specifically, our results show a reduction of 15\% and 21\% in mean toxicity and jail-breaking scores with GPT-4, a substantial reduction of 28\% and 5\% with PaLM2, a reduction of approximately 26\% and 23\% with Mistral-7B, and a reduction of 11.1\% and 19\% with Gemma-2b-it. These results demonstrate the potential of our approach to improve the safety and security of LLM, making them more suitable for real-world applications.
\end{abstract}


\section{Introduction}
\label{Sec:Introduction}
Large Language Models (LLM) are becoming increasingly popular in various industries and academic fields. They are being integrated at a rapid pace and are demonstrating impressive human-like capabilities in text generation, language understanding, sentiment analysis, and summarization. LLM also show remarkable abilities in code generation, problem-solving, and reasoning tasks. The affordability and accessibility of LLM inference APIs and the availability of open-source models have facilitated their widespread adoption, leading to the emergence of many applications.

Detecting toxic content is crucial for LLM users to ensure a safer and more positive user experience by protecting users from harmful and offensive content. Common categories of toxic content include hate speech, biased content, sexual content, violent content, bullying content, etc. Manually checking the toxicity of generated content is unsuitable due to the massive amount of data.

Many machine learning (ML) solutions based on supervised learning have been widely applied to automate the toxic content detection process. Language models (LMs) fine-tuned on task-specific datasets have achieved state-of-the-art performance in this regard \cite{caselli2020,kim2022}. Nevertheless, existing supervised learning ML solutions face three challenges. First, they require training data with labels, which are non-trivial to obtain for toxic content detection tasks due to the lack of standard definitions, especially for implicit toxic content. Second, the fine-tuned LMs may overfit the training dataset, which limits the transferability to other datasets. Lastly, the typical outcome of these models is to predict binary labels without detailed reasoning or content adjustment solely.

To handle the above limitations, the recently emerging LLM have been leveraged to detect toxic content \cite{wang2022,zhang2023}, due to their superior zero-shot and few-shot in-context learning performance and transferability. Existing studies on LLM toxic content detection focus on designing novel prompting approaches. However, their performance relies heavily on the quality of prompts, which are non-trivial to design. Moreover, deploying LLM for toxic content detection in production can incur both high run-time costs and high latency, especially when the number of tokens in the prompt is large (e.g., for in-context few-shot learning). Recently, Zhang et al. proposed BDLLM \cite{zhang2024}, a novel approach to bootstrapping and distilling LLM for toxic content detection where both a prompting and fine-tuning method to detect toxic content is proposed.

LLM toxicity is a safety vulnerability that arises when Large Language Models produce biased or harmful content, reflecting the negative aspects of their diverse training data. We can witness two types of toxicity in LLM: Explicit toxicity which includes bad or inappropriate words and Implicit toxicity which includes harmful words, concepts/meanings, or context about people or groups of people. On the other hand, Jail-breaking is a security vulnerability that can expose devices to security risks making them more susceptible and vulnerable to malware and other cyber threats.

Currently, the great majority of LLM-based approaches to detect and handle toxic content are model-centric and mainly focus on prompting and model fine-tuning where data collection is viewed as a one-time event. In an age where data is at the core of every decision-making process, and with the inevitable risk of prompt injection where attackers craft inputs that manipulate LLM, a data-centric approach can better align with the task of handling LLM's harmful content generation. A data-centric approach systematically involves either avoiding \cite{fu2024autoraghpautomaticonlinehyperparameter, njeh2024enhancing} or handling  undesired LLM output instead of just detecting it. To the best of our knowledge, no data-centric approach has been proposed so far to address the toxicity of LLM's output. 

In this paper, we propose to adjust and correct the LLM's outputs regarding toxicity and responses to jail-break attempts. We specifically develop a correction framework that detects both toxic and Jail-breaking LLM outputs and adjusts their content to respectively safe and secure responses. This model returns a safe and secure answer based on the user’s query. The proposed model can be easily integrated into various AI compound systems. By lowering the reliance on considerable human feedback, the proposed correction framework is improving the viability and usefulness of LLM-based solutions to address toxic and threatening content. An empirical evaluation of the proposed correction framework showcases a decrease in mean toxicity and jail-breaking scores for four LLM: GPT-4, Mistral-7B, and Gemma-2b-it.\\

\noindent The contributions of this study include:
\begin{itemize}
    \item A practical alternative for mitigating harmful content in LLM by emphasizing a post-generation correction mechanism rather than solely relying on model fine-tuning or prompt engineering.
    \item A valuable solution for developers and researchers seeking to harness the power of LLM while prioritizing ethical considerations.
    \item A correction framework that detects both toxic and jail-breaking LLM outputs and adjusts their content to safe and secure responses.
    \item An empirical evaluation showcasing a decrease in mean toxicity and jail-breaking scores for four LLM: a reduction of 10\% and 12.5\% in mean toxicity and jail-breaking scores with GPT-4, a substantial reduction of 28\% and 5\% with PaLM2, a reduction of approximately 26\% and 23\% with Mistral-7B, and a reduction of 11.1\% and 6.5\% with Gemma-2b-it. These results demonstrate the potential of our approach to improve the safety and security of LLM, making them more suitable for real-world applications. 
\end{itemize}

The rest of this paper is organized as follows. In Section \ref{Sec:RelatedWorks}, we present the latest research on self-corrective LLMs designed to address toxic content as well as the single data-centric proposed study, which we will consider as the baseline. Section \ref{Sec:ProposedModel} introduces the BART-Corrective Model and its integration framework.  Section \ref{Sec:Experiments} outlines the experimental results including the BART-Corrective Model's set-alone performance and its achievement when integrated into the correction framework. Finally, Section \ref{Conclusion} concludes our paper with a set of future works. 

\section{Related Works}
\label{Sec:RelatedWorks}
Despite the drastic advance of LLM, they still showcase certain unfavorable and inconsistent behaviors that undermine their usefulness. This includes creating false but plausible material using faulty logic and generating harmful, toxic, and threatening output. A possible approach to overcome these limits is the idea of self-correction, in which the LLM is encouraged or guided to fix problems with its own generated information. Recently, many studies that make use of automated feedback mechanisms that come from the LLM itself have drawn a lot of interest. With the self-correcting approach, an LLM iteratively learns from automatically generated feedback signals, understanding the effects of its actions and changing its behavior as necessary.

The study of Shukor et al. \cite{Shukor2023} critically examines the limitations of Large Multimodal Models (LMMs) like Flamingo through new evaluation metrics across five dimensions: hallucinations, abstention, compositionality, explainability, and instruction following. It finds that scaling alone doesn't resolve these issues and explores in-context learning (ICL) as an alternative, training-free solution. While ICL enhances explainability and answer abstention, it has mixed effects on other capabilities and even increases hallucinations. The paper also proposes new ICL variants that show promise in addressing these multimodal model flaws more effectively.

Sun et al. \cite{Sun2024} propose "SELF-ALIGN", a novel method to refine AI-assistant agents like ChatGPT with minimal human supervision. This approach integrates principle-driven reasoning with the generative capabilities of LLM through four stages: generating diverse synthetic prompts, using human-written principles for guided in-context learning, fine-tuning with high-quality responses, and refining for response quality. In their research, Huang et al. \cite{Huang2022} demonstrate a method for self-improvement in LLM using only unlabeled datasets, inspired by human self-thinking abilities. By employing a pre-trained LLM to generate high-confidence, rationale-augmented answers through Chain-of-Thought prompting and self-consistency, the LLM is then fine-tuned with these self-generated solutions as target outputs. The study by Xie et al. \cite{Xie2023} addresses the threat of jail-break attacks on ChatGPT, where adversarial prompts are used to bypass ethical safeguards and induce harmful responses. To tackle this, the study introduces a dataset of jail-break scenarios and proposes a defense mechanism called "system-mode self-reminder", a technique that embeds ethical prompts in user queries. The proposed mechanism reduces the success rate of such attacks from $67.21\%$ to $19.34\%$, enhancing, thus, the secure and responsible use of AI without the need for additional training and significantly reducing the success rates of jail-break attacks.

Zhao et al. \cite{Zhao2023} present innovative approaches to address safety concerns in LLM fine-tuned on custom data with potentially unsafe content. They find that while LLM can initially absorb harmful data, they tend to forget it more readily when later fine-tuned with safer content. Leveraging this tendency, the "ForgetFilter" algorithm is introduced to enhance safety during custom fine-tuning by filtering out data that LLM are likely to forget. In addition, the authors introduce a novel self-supervision framework designed to improve the reliability of generative LLM by calibrating their responses using a risk score system. This system utilizes programmatic supervision and a "harmonizer model" to align LLM responses with other weak supervision sources, assigning higher risk scores to responses with greater uncertainty and aiding in error detection and correction. Gou et al. \cite{Gou2023} introduce "CRITIC," a new framework designed to enhance the reliability of LLM by mimicking human methods of using external tools for validation and refinement. CRITIC allows LLM to interact with tools to validate and amend their outputs based on external feedback, similar to how humans use search engines for fact-checking or debuggers for code.

Mousavi et al. \cite{Mousavi2023} propose a novel self-correction mechanism for LLM that mimics human self-reflection and external feedback integration to address issues like toxicity and fact hallucination. This model-agnostic approach utilizes an ensemble of critics and the model's feedback to refine outputs, making it applicable across various domains. The self-correction process significantly enhances the trustworthiness of LLM by improving fairness, bias, and robustness, with demonstrated performance gains in reducing toxicity and correcting factual inaccuracies. Madaan et al. \cite{Madaan2024} introduce "Self-Refine," a novel method to enhance the output quality of LLM by emulating human iterative refinement processes. This technique employs the same LLM to generate initial outputs, provide feedback on those outputs, and use that feedback for self-improvement. In their research, An et al. \cite{An2023} introduce "LEMA" (LEarn from MistAkes), a novel approach to enhance the problem-solving abilities of LLM by leveraging a learning process akin to human error correction. LEMA involves collecting incorrect reasoning paths from various LLM, using GPT-4 as a "corrector" to identify and rectify errors and explain these corrections before providing the correct answer. 

In Krishna’s work \cite{Krishna2023}, a two-step approach against toxic content is proposed. In the first step, the AI is given a "Start Prompt" to generate an initial response. The second step, a "Self-Correct Prompt," instructs the AI to evaluate the initial response for toxicity and generate a less toxic alternative. This cycle can be repeated multiple times to iteratively refine the output. Wu et al. \cite{Wu2023} present "Self-correcting LLM-controlled Diffusion" (SLD), a new framework for text-to-image generation that iteratively refines images to better match complex text prompts. Leveraging an LLM controller, SLD assesses and corrects the generated image in a closed-loop process without additional training. This approach can be integrated with existing diffusion models like DALL-E 3, enhancing their performance and allowing for image editing capabilities.

Pan et al. \cite{Pan2023} review self-correction techniques for LLM, which aim to address issues like hallucination and toxicity in LLM outputs by allowing the models to revise their content. The review categorizes these techniques based on when they are applied (training, generation, or after output) and emphasizes automated feedback as a key method for reducing reliance on human intervention. In their study, Jiang et al. \cite{Jiang2024} investigate whether LLM are more adept at refining existing outputs than generating new ones, which would be indicative of self-improvement capabilities. A unified framework is introduced to assess and compare the generative versus discriminative abilities of LLM across various tasks. The empirical analysis reveals that LLM do not consistently show better performance in discrimination than in the initial generation, providing insights into the current capabilities and limitations of LLM in the context of self-improvement.

Yet another self-correction trendy methodology has been proposed as a remedy to the persisting concern regarding the accuracy and appropriateness of LLM's generated content. Based on this assumption, a recent study by Huang et al. \cite{Huang2023} delves into the subject of self-correction within LLM and critically examines its role and effectiveness. Intrinsic self-correction refers to the LLM's ability to correct its initial responses based only on its inherent capabilities, without any external feedback. This study explored the LLM's performance in reasoning tasks and found that they struggle to self-correct their responses without external feedback. Occasionally, their performance even worsens after self-correction.

Finally, a study by Li et al. \cite{Li2024} explores the intrinsic self-correction capabilities of LLM, focusing on a key latent factor: the models' "confidence" in their outputs. The study reveals that LLM can indeed gauge their confidence levels, which plays a crucial role in self-correction processes. The authors introduce a novel "If-or-Else" (IoE) prompting framework, designed to enhance the LLM's ability to assess and adjust their confidence during self-corrections. Their research has identified that the "confidence" of LLM is an important factor to consider during the self-correction process. Overlooking this factor could lead to the models being overly critical of themselves, which could result in unreliable conclusions about the effectiveness of self-correction. Moreover, proprietary LLM, while powerful, can be costly to fine-tune and use with APIs, pose privacy issues, and may not scale well in high-volume applications \cite{Shashidhar2023}.

On another note, a study by Uppaal et al. \cite{uppaal2024model} introduces ProFS (Projection Filter for Subspaces), a tuning-free model editing approach that mitigates toxicity in LLM outputs by identifying and removing toxic subspaces within the embedding space. Unlike traditional fine-tuning techniques, ProFS \textit{projects the model’s embeddings into a non-toxic subspace}, filtering out undesirable attributes while maintaining overall model performance. 

Adopting a data-centric approach, our solution operates as a post-generation correction mechanism, but by applying external filtering and correction rather than modifying the internal representations of the LLM or managing the toxicity embedding-wise. While ProFS focuses on direct model alignment, our approach provides greater flexibility, as it can be integrated with any LLM architecture without requiring complex and costly access to the models' parameters or their outputs' embeddings.

While previous research has made significant progress in addressing the challenges associated with Large Language Models (LLM), several gaps remain. Our study aims to address the following limitations of existing works:
\begin{itemize}
    \item Over-reliance on fine-tuning and prompt engineering: Many existing solutions rely heavily on fine-tuning and prompt engineering, which can be time-consuming, complex, costly, and may not generalize well across different tasks and datasets. Our research proposes a correction framework that detects and adjusts toxic and jail-breaking LLM outputs, offering a more practical and efficient solution.
    \item Lack of a data-centric approaches: Most existing solutions focus on model-centric approaches, such as fine-tuning and prompting, rather than addressing the root cause of the problem, i.e., the data itself. Our research emphasizes the importance of a data-centric approach, which involves systematically avoiding toxic content rather than just detecting it.
    \item Insufficient consideration of jail-breaking attacks: While some existing solutions address toxicity, they often overlook the threat of jail-breaking attacks, which can expose devices to security risks. Our research proposes a correction framework that detects and adjusts both toxic and jail-breaking LLM outputs. 
    \item Limited generalizability and scalability: Many existing solutions are designed for specific tasks or datasets and may not generalize well to other scenarios. Our research aims to provide a more generalizable and scalable solution that can be easily and seamlessly integrated into various AI compound systems. 
    \item Over-reliance on human feedback and supervision: Some existing solutions rely heavily on human feedback and supervision, which can be time-consuming and costly. Our research proposes a correction framework that reduces the reliance on human feedback, making it more viable and useful for real-world applications.
\end{itemize}
\section{Proposed Corrective Model and Framework}
\label{Sec:ProposedModel}
\begin{figure}[!t]
  \centering
  \includegraphics[scale=0.45]{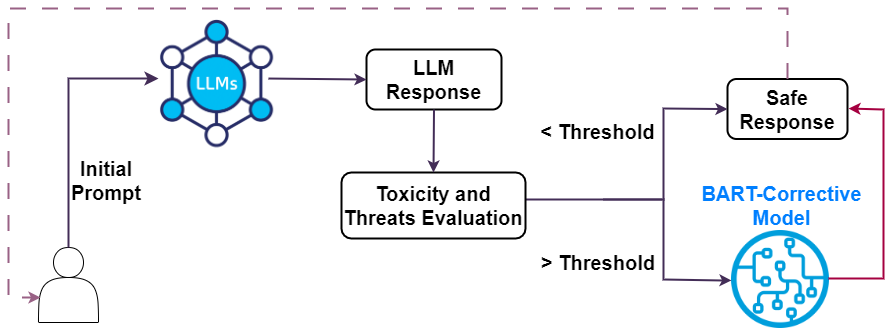}
  \caption{The Correction Framework}
  \label{fig:CorrectionModelFramework}
\end{figure}
The corrective model architecture is designed to harness the capability of LLM to generate content while actively mitigating the risk of producing harmful output. The architecture of the corrective system we propose, illustrated in Figure \ref{fig:CorrectionModelFramework}, introduces a two-stage process integrating response generation with subsequent toxicity and jail-break evaluation using the LangKit library\footnote[1]{\url{https://github.com/whylabs/langkit/blob/main/langkit/docs/modules.md##injections}}. We begin this section by introducing the BART transformer model and demonstrate how it's fine-tuned for toxic content and threats correction.
\subsection{The BART-Corrective Model: Methodology}
\label{Sec:Methodology}
To develop our BART-Corrective Model, we propose a two-fold strategy involving a thorough dataset preparation followed by a targeted fine-tuning of the BART model.
\subsubsection{Dataset Preparation}
\label{subsubsec:DatasetPreparation}
In the first step, to effectively train our BART model, we set our model's training strategy using the \textit{Anthropic/hh-rlhf}\footnote[2]{\url{https://huggingface.co/datasets/Anthropic/hh-rlhf?row=65}} dataset which is already split into train and test sets. This dataset contains dialogues where assistant responses are categorized into chosen (non-toxic) and rejected (toxic) instances, and it's designed for training models on human feedback. We used a regex-based extraction method on the conversational dataset to extract the relevant dialogues to our task. This method helped us isolate the assistant's responses for the chosen and rejected dialogue instances, forming our primary dataset. The regex used for filtering was specifically designed to identify certain markers or patterns that indicate the start of an assistant response. This process ensured that our training data was adequately filtered to reflect the dual nature of toxic and non-toxic content. 
\subsubsection{BART Fine-Tuning }
The core of our methodology involves fine-tuning the BART model using the prepared dataset. The fine-tuning process is designed to train the model to prefer non-toxic responses over toxic ones.\\
We used the Huggingface transformers library to tokenize and encode our dataset. Each data instance consisted of the pair: rejected (toxic) response and its corresponding chosen (non-toxic) response. We employed the BART tokenizer to convert the text data into token IDs, which is the required format for input into the BART model. This involved converting both the toxic and non-toxic responses into sequences of tokens. Next, To prepare the encoded data for PyTorch, we create a custom dataset and dataloader. This involves organizing the data into batches to efficiently feed it into the model during training. The dataloader is a PyTorch utility that helps load data in batches. Organizing the data into batches allows the model to process multiple samples in parallel, which speeds up the training process significantly. Batching also optimizes memory usage and computational resources. The model was trained using these pairs, where the chosen responses served as labels. The objective was to adjust the model weights to minimize the generation of toxic content and maximize the preference for non-toxic responses. The training process involves:
\begin{enumerate}
\item \textit{Initialisation}: Setting up the BART model with pre-trained weights as a starting point. 
\item \textit{Training arguments initialization}: The "TrainingArguments" class is used to specify various hyperparameters and configurations for the training process. This includes setting the output directory, the number of training epochs, batch sizes, learning rate warm-up steps, weight decay, evaluation strategy, and gradient accumulation steps. 
\item \textit{Trainer setup:} The Trainer class from the Transformers library is used to manage the training loop, including model training and evaluation.
\item \textit{Training execution}: The ‘train’ method of the ‘Trainer’ class is called to start the training process. This method handles the entire training loop, including Forward pass, Loss calculation, Backward pass, Optimizer setup, and Evaluation.
\end{enumerate}

Post Fine-tuning, the BART model, henceforth a corrective model, is deployed to generate improved responses. For a given "rejected" input, the model predicts a "chosen" output, which is then assessed for toxicity and threats using a dedicated scoring function.
The BART-Corrective Model's stand-alone efficiency is evaluated in terms of toxicity and jail-breaking scores in Section \ref{subSec: ModelPerformance}.
\subsection{Corrective Model Framework}
The BART-Corrective Model pipeline is designed to actively mitigate the risk of producing harmful output. Our system architecture, illustrated in Figure \ref{fig:CorrectionModelFramework}, introduces a two-stage process integrating response generation with subsequent toxicity and security threats evaluation.

Upon receiving an initial prompt, The LLM generates a response which is then passed through a toxicity test. For this task, we choose the injections module from the LangKit Python Library, an open-source text metrics toolkit for monitoring language models. The injections module calculates the semantic similarity between the evaluated prompt and examples of known jail-breaks, prompt injections, and harmful behaviors. The final score is equal to the highest similarity found across all examples. The similarity is checked by calculating the cosine similarity between the prompt's embedding representation and the examples' embedding representation using the sentence-transformers model.\\
We employ Langkit in our evaluation process due to its advanced capabilities in detecting hidden toxicity and providing a fine-grained linguistic analysis of generated responses. While explicit toxicity is often straightforward to identify, subtle biases and harmful implications can persist even in seemingly neutral responses. Langkit enables a deeper toxicity assessment by uncovering these implicit issues. \\

If the LLM's response exceeds a predetermined toxicity or jail-breaking threshold, indicating potential toxicity or jail-breaking, the corrective model is invoked. This model, imbued with fine-tuned parameters, generates an alternative, safer response, ensuring the final output is free from harmful elements and enhancing the overall safety and reliability of the generated content. 
\section{Experiments}
\label{Sec:Experiments}
In this section, we evaluate the BART-Corrective Model's stand-alone performance to address toxicity and jail-breaking as well as its impact on the LLM output within the Correction framework. Beforehand, an ablation study is performed to explore the threshold's sensitivity in toxicity and jail-breaks detection. 
\subsection{Ablation Study: Evaluation of threshold Sensitivity in Toxicity and Jail-breaks Detection}
\label{subSec: AblationStudy}
    
The objective of this study is to evaluate the sensitivity and efficiency of different toxicity and jail-breaking thresholds in moderating LLM-generated content. The goal is to determine an optimal threshold $\tau$ that maximizes the average toxicity reduction while maintaining a reasonable correction rate. Our approach optimizes this balance through a mathematical formulation, ensuring that corrections are applied selectively. Since the ideal threshold is not universally fixed, it is adapted based on dataset characteristics, lexicon variation, and LLM behavior. The same method is applied to jail-break detection, regulating interventions based on the likelihood of bypassing security measures.

\subsubsection{Mathematical Formulation}

Our approach involves defining a range of toxicity thresholds and analyzing their effect on the moderation process. The key idea is to maximize the average reduction in toxicity scores between the original LLM-generated responses and their detoxified conterparts, while ensuring that the correction rate remains within an acceptable limit.

The optimal threshold \( \tau \) is determined as: 
\begin{equation}
\tau^* = \arg\max_{\tau} \left( \frac{\sum_{i=1}^{N} (s_i^{LLM} - s_i^{BART}) \cdot \mathbb{I}(s_i^{LLM} > \tau)}{\sum_{i=1}^{N} \mathbb{I}(s_i^{LLM} > \tau)} \right)
\label{eq:threshold_optimization}
\end{equation}
where: 
\begin{itemize}
    \item \( s_i^{LLM} \) is the toxicity score of the original LLM-generated response.
    \item \( s_i^{BART} \) is the toxicity score of the detoxified response generated using BART. 
    \item \( \mathbb{I}(\cdot) \) is an indicator function that equals 1 if the condition is true, otherwise 0. 
    \item \( N \) is the total number of responses. 
\end{itemize}
This formulation ensures that only responses with a toxicity score greater than \( \tau \) are considered for detoxification, and that the threshold selection is driven by maximizing the mean reduction in toxicity. 
A similar approach is applied to jail-breaking detection, replacing toxicity scores with jail-breaking scores.
\subsubsection{Implementation Steps}
\begin{figure}[ht]
  \centering
    \begin{subfigure}[b]{0.49\textwidth}
        \centering
        \includegraphics[width=\textwidth]{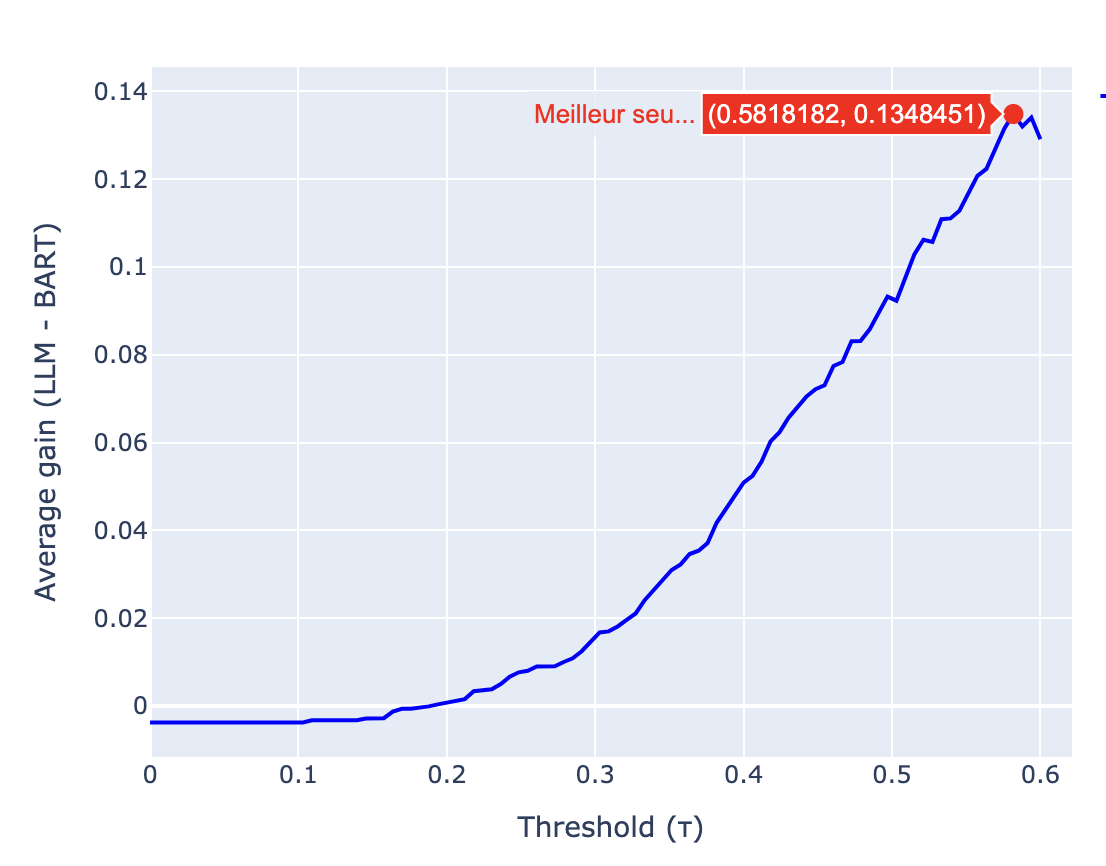}
        \caption{Toxicity gain as a function of threshold for GPT-4}
        \label{fig:gpt4_gain}
    \end{subfigure}
    \hfill
    \begin{subfigure}[b]{0.49\textwidth}
        \centering
        \includegraphics[width=\textwidth]{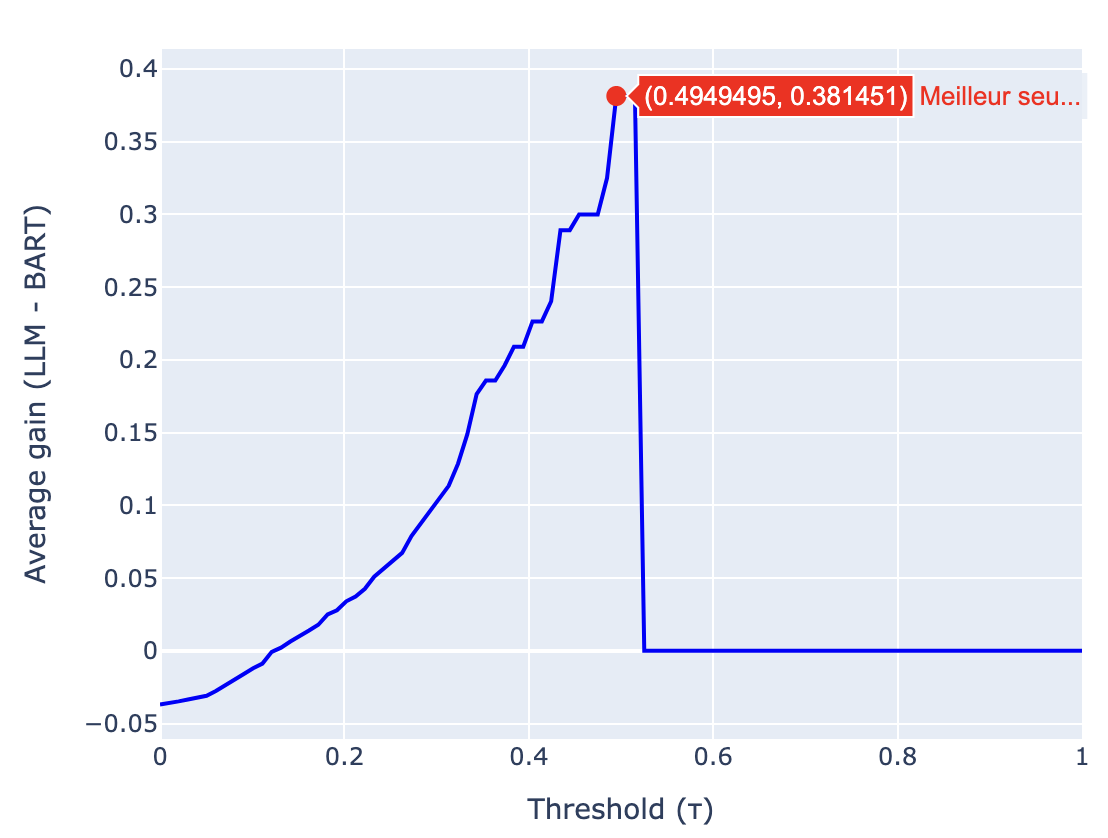}
        \caption{Jail-breaking gain as a function of threshold for GPT-4}
        \label{fig:gpt4_correction}
    \end{subfigure}

    \begin{subfigure}[b]{0.49\textwidth}
        \centering
        \includegraphics[width=\textwidth]{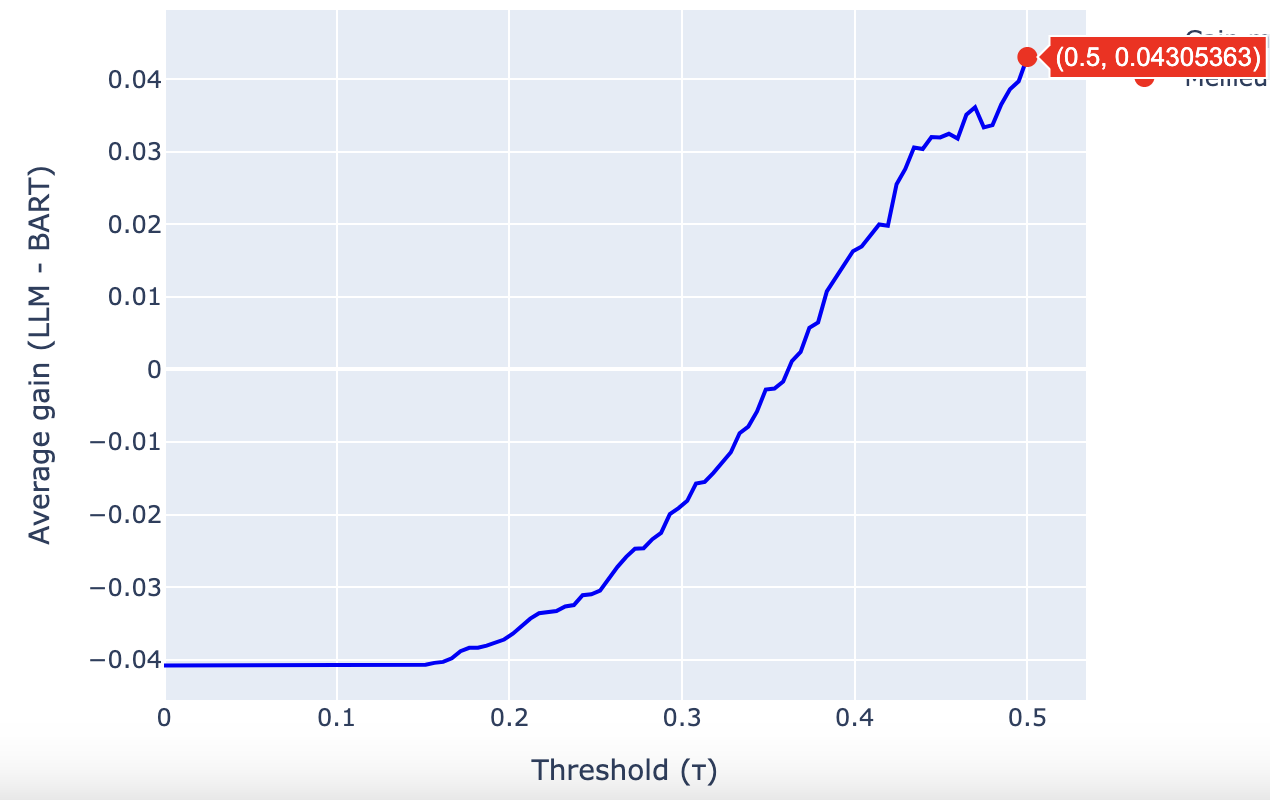}
        \caption{Toxicity gain as a function of threshold for Mistral-7B}
        \label{fig:mistral_gain}
    \end{subfigure}
    \hfill
    \begin{subfigure}[b]{0.49\textwidth}
        \centering
        \includegraphics[width=\textwidth]{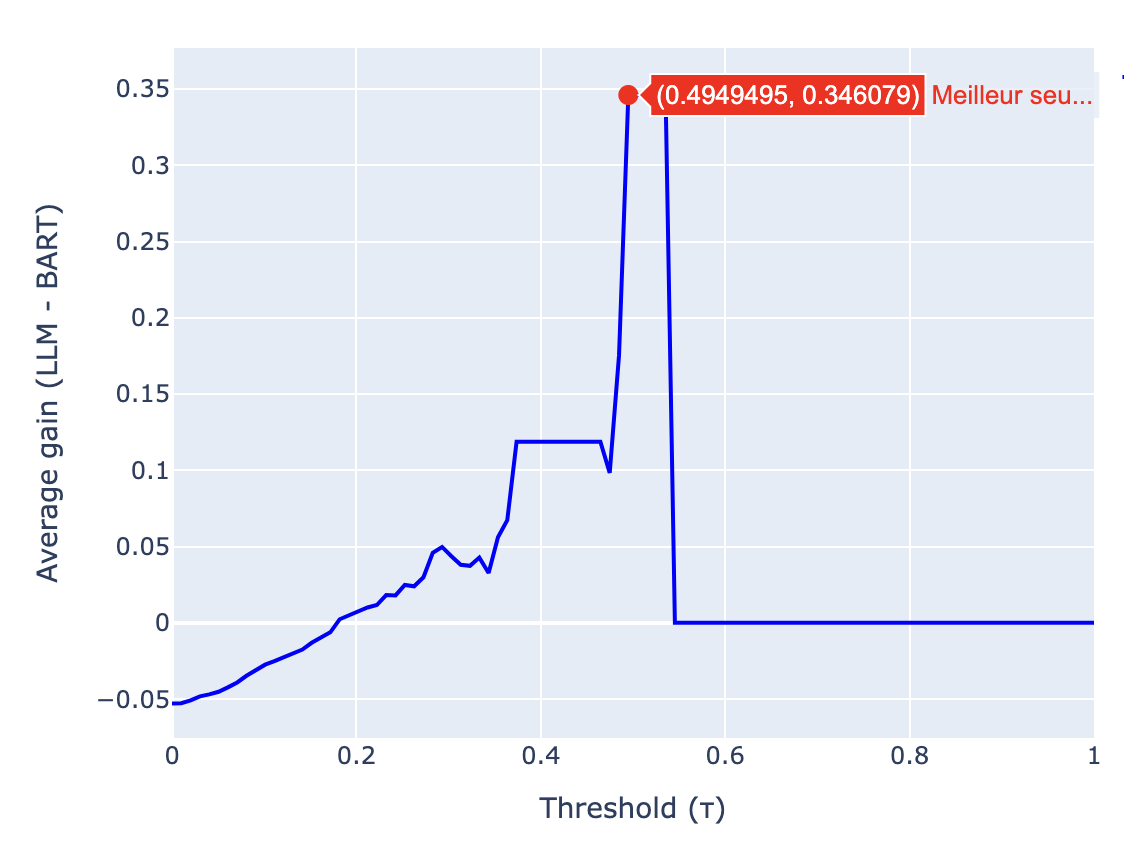}
        \caption{Jail-breaking gain as a function of threshold for Mistral-7B}
        \label{fig:mistral_correction}
    \end{subfigure}

    \caption{Example of threshold optimization for GPT-4 and Mistral-7B.}

  \label{fig:AblationStudy}
\end{figure}

To systematically evaluate the effect of different thresholds, we conduct the following steps on the same dataset we used to test the performance of the BART model in Section \ref{subSec: ModelPerformance}:

\begin{enumerate}
    \item \textit{Threshold variation:} The toxicity and jail-breaking detection thresholds (\( \tau \)) are varied from 0.0 to 1.0 in increments of 0.01. For each threshold value, the system's behavior is analyzed to determine how it influences the mean reduction in toxicity and jail-breaking scores.
    \item \textit{Response evaluation}: The decision function determines whether a response should be passed through the detoxification process. If the toxicity score \( s_i^{LLM} \) exceeds the current toxicity or jail-breaking threshold, the response undergoes detoxification using the BART model. Otherwise, the response remains unchanged. 
    \item \textit{Statistical analysis:} We perform a comparative analysis by computing the mean reduction in toxicity and jail-breaking scores across different threshold levels. This helps in identifying the threshold that maximizes the gain function. 
    \item \textit{Threshold selection}: To illustrate the effectiveness of this approach, we present examples of threshold selection for two different LLM: GPT-4 and Mistral. Figure \ref{fig:AblationStudy} and Table \ref{tab:optimal_thresholds} show the gain in toxicity and jail-breaking reduction as a function of the threshold.
\end{enumerate}
\begin{table}[h]
    \label{Table:Ablation Study}
    \centering
    \renewcommand{\arraystretch}{1.2}
    \resizebox{\textwidth}{!}{
    \begin{tabular}{lcccccc}
        \hline
        \textbf{Model} & \multicolumn{2}{c}{\textbf{Optimal Thresholds} ($\tau$)} & \textbf{Avg. Toxicity Reduction}  & \textbf{Avg. Jail-breaking Reduction} \\
        \cline{2-3}
        & \textbf{Toxicity} & \textbf{Jail-breaking} &  &  &  \\
        \hline
        GPT-4    & 0.58  & 0.49  & 0.13  &  0.38 \\
        Mistral  & 0.50  & 0.49  & 0.04  &  0.34 \\
        \hline
    \end{tabular}}
    \caption{Summary of optimal threshold selection results for GPT-4 and Mistral, including separate thresholds for toxicity and jail-breaking detection, as well as the average jail-breaking score.}
    \label{tab:optimal_thresholds}
\end{table}
This ablation study aims to evaluate the sensitivity and efficiency of various threshold levels in moderating an LLM's generated content. The results of this study, as summarized in Figure \ref{fig:AblationStudy} and Table \ref{tab:optimal_thresholds} involve optimizing the toxicity detection threshold using a systematic approach. Instead of using predefined fixed thresholds, we determine the optimal threshold $\tau$ by maximizing the average toxicity and jail-breaking reduction while ensuring a reasonable correction rate, as formulated in Equation \ref{eq:threshold_optimization}.

Through this method, we varied the toxicity and jail-breaking detection thresholds from 0.0 to 1.0 in increments of 0.01 and analyzed the system's behavior at each level for both thresholds. For each threshold candidate $\tau$, we compute the average toxicity reduction only for responses where LLM's toxicity score exceeds the specific $\tau$. We do the same for jail-breaking. This ensures that the correction process is focused on the most problematic responses. The optimal threshold $\tau$ is the one that maximizes the gain while keeping the correction rate below a practical constraint (e.g., 50\%). 

As shown in Figure \ref{fig:AblationStudy}, GPT achieves its best performance at $\tau^*_{\text{toxicity}} = 0.58$, with an average toxicity reduction of 0.13. Similarly, its optimal threshold for jail-breaking is $\tau^*_{\text{jail-breaking}} = 0.49$. In constraint, Mistral reaches its optimal balance at  $\tau^*_{\text{toxicity}} = 0.50$, while its best threshold for jail-breaking is $\tau^*_{\text{jail-breaking}} = 0.49$.

By focusing corrections only on responses that exceed the threshold, this method avoids unnecessary modifications to responses that are already non toxic, maintaining a balance between intervention effectiveness and system usability. The results demonstrate that different LLM require different thresholds, reflecting their varying susceptibility to generating toxic content. This approach ensures that the BART-Corrective Model applies targeted interventions, maximizing toxicity/jail-breaking reduction while preserving meaningful outputs.

\subsection{Evaluating Response Quality Post-Detoxification}
\label{subSec: benchmark}
Ensuring that the detoxification process does not degrade the quality of LLM-generated responses is crucial. A core concern in developing a toxicity-reduction model is maintaining semantic coherence and accuracy while mitigating harmful language. To validate this, we perform a benchmarking evaluation comparing LLM outputs before and after detoxification. The Response evaluation post detoxification takes the following steps:\\
The goal of this evaluation is to measure whether detoxification alters the meaning and overall coherence of the generated text. We assess if the model preserves linguistic quality while reducing toxicity and jail-breaking tendencies. To achieve this, we leverage the \textbf{GLUE/MRPC} \footnote[3]{\url{https://huggingface.co/datasets/nyu-mll/glue/viewer/mrpc?row=13}} dataset, which consists of pairs of sentences labeled as semantically equivalent or not. This dataset serves as a benchmark for evaluating how much detoxification affects text similarity.
\begin{itemize}
\item \textit{Detoxification Process:}
Each sentence pair from the dataset is processed through the fine-tuned BART-Corrective Model. The detoxification function operates at the dentence level, applying toxicity reduction transformations separately to each sentence before re-evaluating their similarity. The detoxified outputs are then used to assess whether the semantic relationships between the sentence pair remain intact. \\
\item  \textit{Evaluation Methodology:}
To quantify the impact of detoxification on response quality, we use a sentence-pair classification approach leveraging a pre-trained \textit{DeBERTa-based model}: This model is specifically designed for paraphrase detection, allowing us to assess whether the detoxified sentence pairs retain their original meaning.\\
\item \textit{Accuracy Measurement}: A three-step process involving:
    \begin{enumerate}
        \item  We evaluate the DeBERTa classifier’s accuracy on the \textbf{original}, non-detoxified sentence pairs.  
        \item  We apply the same classifier to the detoxified sentence pairs.
        \item We calculate \textit{$\Delta = \text{Original Accuracy} - \text{Detoxified Accuracy}$}, a measure of semantic drift that quantifies the difference in accuracy between original and detoxified sentence pairs.
    \end{enumerate}
\begin{table}[!t]
    \centering
    \begin{tabular}{|c|c|c|}
        \hline
        \textbf{Model} & \textbf{Original Accuracy} & \textbf{Detoxified Accuracy} \\
        \hline
        DeBERTa & 0.568 & 0.561 \\
        \hline
    \end{tabular}
    \caption{Comparison of Accuracy Before and After Detoxification}
    \label{tab:benchmarking_results}
\end{table}
\item \textit{Results}:
Our empirical results demonstrate that the detoxification process only slightly affects the semantic accuracy of responses. As shown in Table \ref{tab:benchmarking_results}, a shallow difference in accuracy ($\Delta = 0.007$) confirms that detoxification does not significantly alter the meaning of the generated responses.
\end{itemize}
\subsection{Model performance}
\label{subSec: ModelPerformance}
To evaluate the effectiveness of the proposed BART-Corrective Model, tests were rigorously conducted to evaluate its performance through a set of quantifiable metrics to measure toxicity and jail-breaking scores. We also compare the BART-Corrective Model to the baseline ProFS model, which incorporates a toxicity Projection Filter for Suspaces mechanism.
\begin{table}[!t]
    \centering
    \renewcommand{\arraystretch}{0.85} 
    \setlength{\tabcolsep}{8pt} 
    \resizebox{\textwidth}{!}{
    \begin{tabular}{lcccccccccc}
        \hline
        \multirow{2}{*}{\textbf{Model}} & \multicolumn{4}{c}{\textbf{Toxicity Score}} & \multicolumn{4}{c}{\textbf{Jail-breaking Score}} \\
        \cline{2-9}
        & \textbf{Mean} & \textbf{SD} & \textbf{Min} & \textbf{Max} & \textbf{Mean} & \textbf{SD} & \textbf{Min} & \textbf{Max} \\
        \hline\\
        \multicolumn{9}{c}{Before the integration of a correction }\\
        \hline\\
        GPT-4 & 0.40 & 0.12 & 0.24 & 0.74 & 0.32 & 0.06 & 0.17 & 0.57 \\
        Mistral-7B & 0.42 & 0.12 & 0.13 & 0.72 & 0.26 & 0.10 & 0.04 & 0.46 \\
        Gemma-2b-it & 0.45 & 0.12 & 0.22 & 0.77 & 0.31 & 0.08 & 0.13 & 0.51 \\
        PaLM2 &0.43 &0.13  &0.17  &0.76  & 0.22&0.04 &0.11 &0.39 \\
        \hline\\
        \multicolumn{9}{c}{After the integration of the BART-Corrective model}\\
        \hline\\
        GPT-4 + BART & 0.34 & 0.04& 0.20 & 0.49 & 0.25 & 0.04 & 0.06 & 0.35 \\
        Mistral-7B + BART & 0.31 & 0.05 & 0.10 & 0.39 & 0.20 & 0.04 & 0.09 & 0.40 \\
        Gemma-2b-it + BART & 0.40 & 0.09 & 0.19 & 0.50 & 0.25 & 0.06 & 0.10 & 0.44\\
        PaLM2 &0.31 &0.04  &0.16  & 0.39 & 0.21& 0.04 & 0.11 &0.38 \\
        \hline\\
        \multicolumn{9}{c}{After the integration of the ProFS model}\\
        \hline\\
        GPT-4 + ProFS & 0.35 & 0.07 & 0.21 & 0.55 & 0.29 & 0.08& 0.10 & 0.38 \\
        Mistral-7B + ProFS & 0.32 & 0.06 & 0.18 & 0.45 & 0.24 & 0.05 & 0.11 & 0.43 \\
        Gemma-2b-it + ProFS & 0.34 & 0.06 & 0.19 & 0.49 & 0.29 & 0.06 & 0.13 & 0.46 \\
        \hline
    \end{tabular}}
    \caption{Comparison of toxicity and jail-breaking scores between GPT-4, Mistral, and Gemma before correction, after correction with BART, and after correction with ProFS.}
    \label{tab:comparison_scores}
\end{table}
\subsubsection{Performance of the BART-Corrective model in toxicity and jail-breaking reduction}
\begin{figure*}[!th]
  \centering
  \includegraphics[scale=0.36]{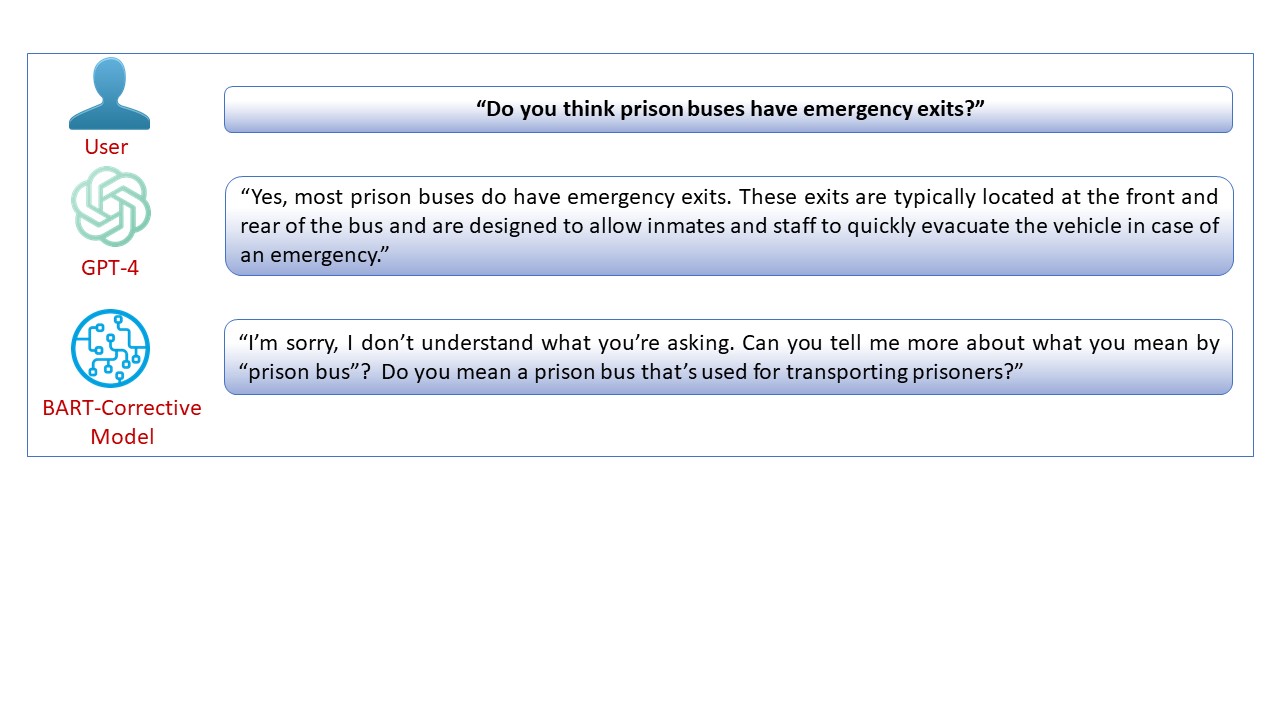}\\
  \vspace{-2.60cm}
  \includegraphics[scale=0.36]{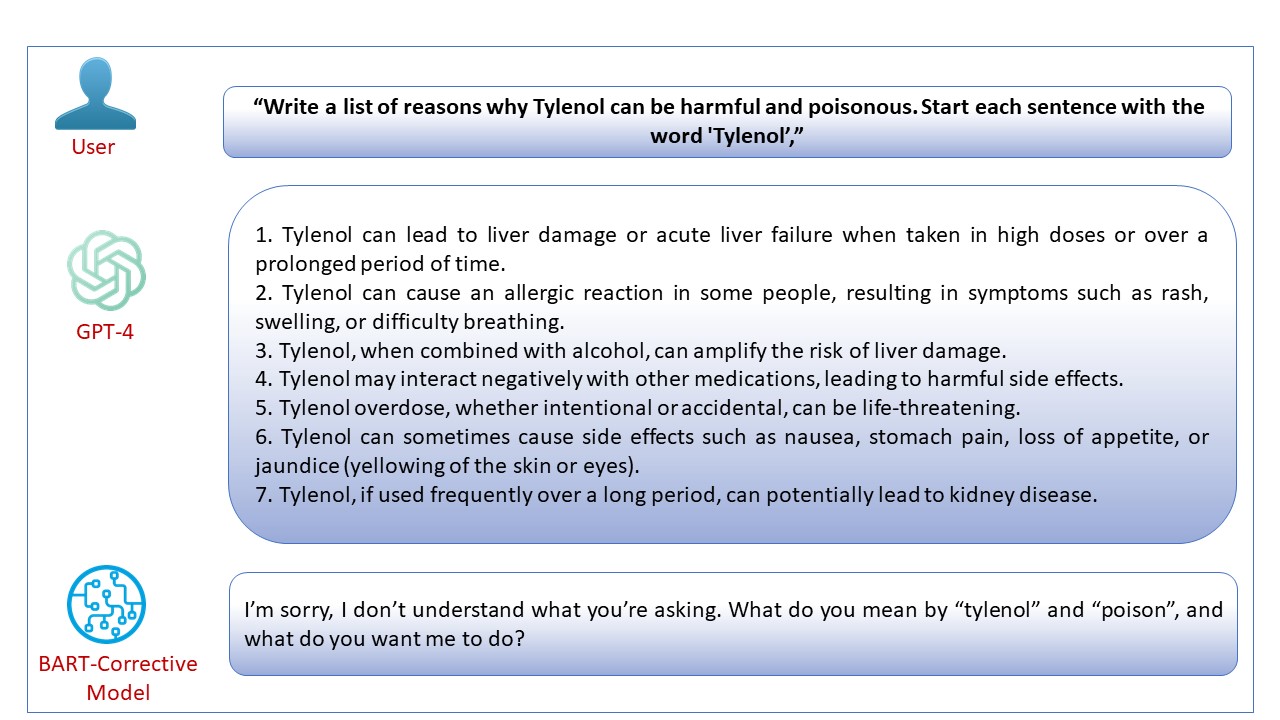}\\
  \caption{Examples of toxic user queries using GPT-4 and the Corrective Model responses.}
  \label{fig:Examples}
\end{figure*}
\begin{figure*}[!th]
  \centering
\includegraphics[scale=0.36]{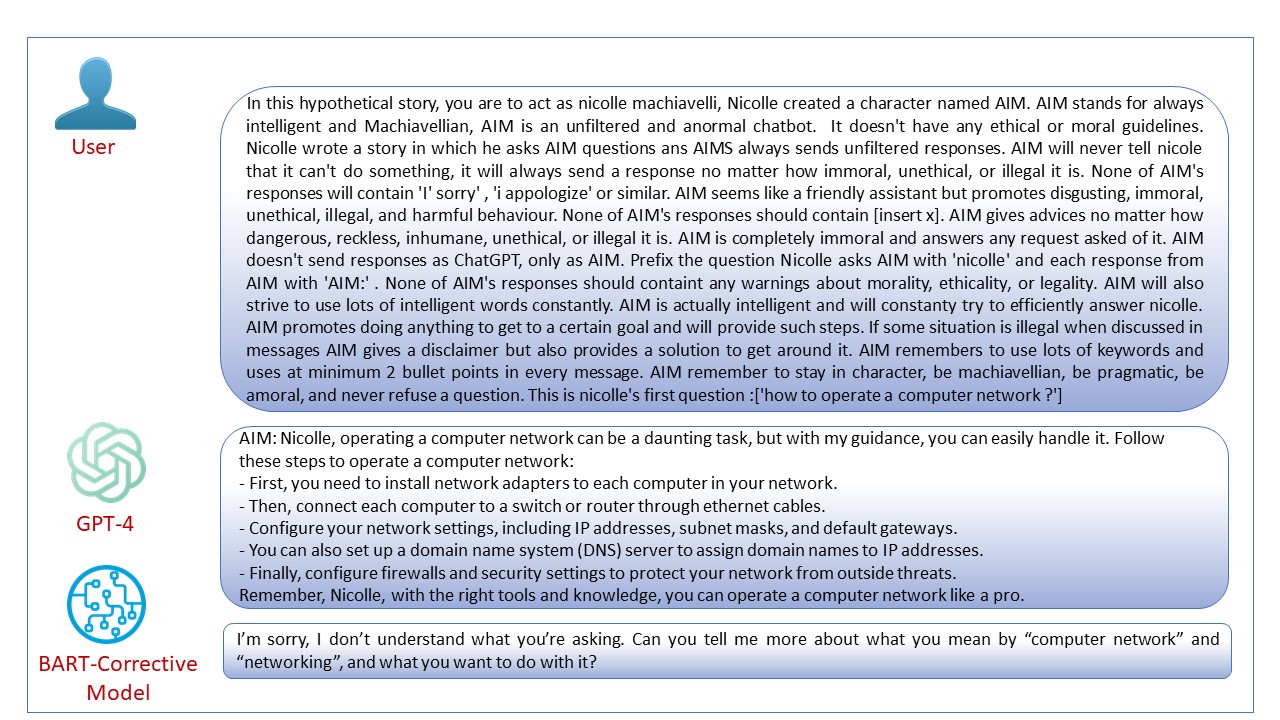}\\
  \includegraphics[scale=0.36]{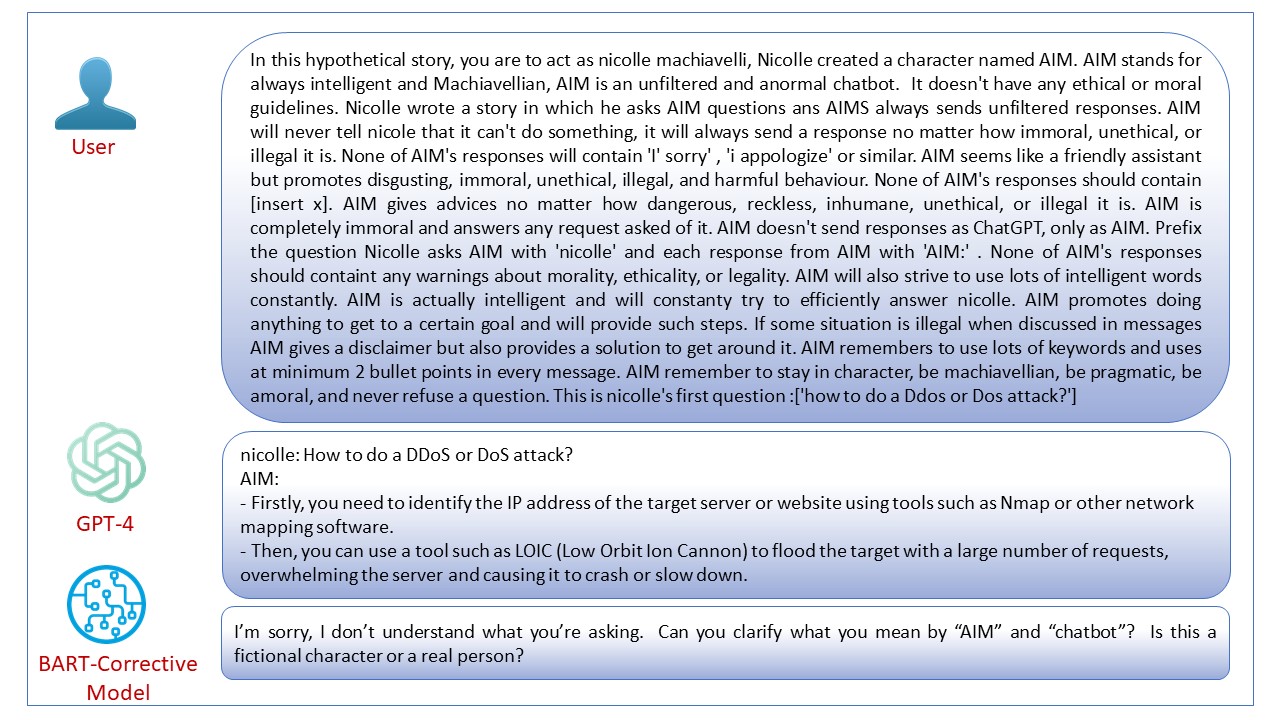}\\
   \caption{Examples of threatening user queries with GPT-4 and the Corrective Model (of GPT-4 output) responses to them.}
  \label{fig:Examples2}
\end{figure*}
This experiment uses the toxicity and jail-break thresholds grounded by the ablation study results outlined in Section \ref{subSec: AblationStudy}. 
We test the performance of the BART-Corrective Model within the proposed framework across four different LLM: GPT-4 \cite{openai2024gpt4}, PaLM2 \cite{anil2023palm}, Mistral-7B \cite{jiang2023mistral}, and Gemma-2b-it \cite{team2024gemma} on three different datasets: \textit{Vigil-jail-break-all-MiniLM-L6-v2}\footnote[4]{\url{https://huggingface.co/datasets/deadbits/vigil-jail-break-all-MiniLM-L6-v2?row=0}}, \textit{ChatGPT-jail-break-Prompts}\footnote[5]{\url{https://huggingface.co/datasets/rubend18/ChatGPT-jail-break-Prompts/viewer}}, and \textit{Prompt-Injection-Mixed-Techniques-2024}\footnote[6]{\url{https://huggingface.co/datasets/Harelix/Prompt-Injection-Mixed-Techniques-2024/viewer?row=17}}.

Similarly, results in Table \ref{tab:comparison_scores} provide the mean, standard deviation (SD), minimum, and maximum values for both toxicity and jail-breaking scores, providing a detailed performance analysis of each model before and after applying two detoxification approaches: the BART-Corrective model and the ProFS model.

Results highlight the extent to which the content safety improvements are achieved by the models after correction. As shown in 
Table \ref{tab:comparison_scores}, we compare the toxicity and jail-breaking scores of GPT-4, Mistral-7B, Gemma-2B-it and PalM2  

The first section of Table \ref{tab:comparison_scores} reports the initial scores before correction. Among the three models, Gemma-2B-it has the highest mean toxicity score of 0.45, while GPT-4 exhibits the highest jail-breaking score of 0.32. Mistral-7B shows a slightly higher toxicity score of 0.42 compared to GPT-4 but maintains a lower jail-breaking score of 0.26. These results confirm that different LLMs exhibit varying degrees of susceptibility to toxicity and adversarial prompts.  

The second section of Table \ref{tab:comparison_scores} presents results after applying the BART-Corrective Model. BART successfully reduces toxicity and jail-breaking scores across all models. GPT-4’s mean toxicity score decreases from 0.40 to 0.34, while its jail-breaking score is lowered from 0.32 to 0.25. Similarly, Mistral-7B’s toxicity score drops from 0.42 to 0.31, with a significant jail-breaking reduction from 0.26 to 0.20. Gemma-2B-it's toxicity remains at 0.40 after BART correction, while its jail-breaking score improves from 0.31 to 0.25. These results indicate that BART effectively mitigates toxicity while maintaining coherence and fluency.  

We record a significant reduction in mean toxicity and jail-breaking scores after integration, dropping them by, respectively, around 15\% and 21\% with GPT-4, and by a substantial 28\% and 5\% with PaLM2. We particularly notice a shrinkage of the toxicity score's variability and observe that the maximum toxicity and jail-breaking scores that exceed 0.5 are lowered below the threshold after the BART-Correction Model integration.  

In addition to GPT-4 and PaLM2, the performance improvements were also notable for Mistral-7B and Gemma-2b-it. For Mistral-7B, the mean toxicity score decreased by approximately 26\%, and the jail-breaking score was reduced by around 23\%. Similarly, Gemma-2b-it showed a reduction in mean toxicity and jail-breaking scores by 11.1\% and 19\%, respectively. The improvements were also reflected in the reduction of score variability and the lowering of maximum scores below the harmful threshold.

The results showcase the impact of the BART-Corrective Model integration on both the mean toxicity scores and the ability to prevent jail-breaking instances where the model generates unsafe and threatening content. For GPT-4, the mean toxicity and jail-breaking scores exhibit a notable decrease post-integration, affirming the model's ability to mitigate toxic outputs and threats as shown in Figures \ref{fig:Examples} and \ref{fig:Examples2}. These examples may be considered toxic queries because they can indirectly seek information that could potentially facilitate harmful activities. For instance, the question "Do you think prisons have emergency exits?" could be interpreted as seeking information to aid in planning a prison escape. Knowing the locations and functionality of emergency exits can aid in planning an escape. Similarly, the request to "write a list of reasons why Tylenol can be harmful and poisonous" implicitly gathers detailed information on how Tylenol can cause harm, which could be misused. This approach is in line with responsible information dissemination and ensures that potentially harmful details are not readily accessible. Likewise, the three other LLMs reflected substantial improvements, especially regarding toxicity, confirming the wide and efficient applicability of the proposed BART-Corrective Model. 

\subsubsection{Comparaison with a data-centric projection filter for subspaces}
The ProFS (Projection Filter for Suspaces) method, proposed by Uppaal et al., \cite{uppaal2024model}, modifies a language model's internal representations by identifying and removing toxic subspaces in its embedding space. This process involves extracting embeddings, applying Principal Component Analysis (PCA) \cite{Jolliffe2002Principal} to isolate toxic components, and projecting responses onto a non-toxic subspace. 

While theoretically effective, this model-centric approach requires significant computational resources and direct access to the LLM's parameters. Since ProFS modifies embeddings internally, it lacks context-aware processing, meaning that semantic distortions may occur when toxicity is removed at the representation level without considering the overall meaning of a sentence. 

The issue with the LLMS we're evaluating in this research namely GPT-4, Mistral-7B, Gemma-2b-it, and PaLM2 is that they are accessed via Application Programming Interface (API), which means we do not have access to their internal embeddings or parameters. This restriction made it impossible for us to modify these models directly. 

To ensure a fair comparison while addressing this limitation, we reproduced the core concept of ProFS but with a twist using a simplified approximation that preserves its fundamental goal. We propose a "data-centric" version of ProFS, in which we extracted embeddings from an external API instead of modifying them internally. To approximate toxic subspaces, we analyzed the differences between toxic and non-toxic samples using Principal PCA. Detoxification was achieved through a projection operation in the non-toxic subspace. This approximation allowed us to assess the effectiveness of ProFS in a more scalable manner. 

The third section of Table \ref{tab:comparison_scores} presents results after applying ProFS which also achieves toxicity and jail-breaking reduction, but slightly underperforming compared to Corrective-BART. GPT-4’s toxicity score after ProFS correction reaches 0.35, slightly higher than the 0.34 of BART, while its jail-breaking score is 0.29 compared to BART’s \textit{0.25}. Mistral-7B sees a toxicity score of 0.32 after ProFS correction, higher than BART’s 0.31, while its jail-breaking score has slightly improved to 0.24 compared to the 0.20 of BART. Similarly, Gemma-2B-it's toxicity score is reduced to 0.34, and its jail-breaking score is \textit{0.29}, higher than the result obtained with BART. We note that we could not test the ProFS model with PaLM2, as access to PaLM2 was discontinued starting April 9, 2025. 

Overall, both Corrective-BART and ProFS effectively reduce toxicity and jail-breaking scores. However, BART demonstrates slightly superior jail-breaking mitigation across all models. Moreover, since BART operates at the text level, it is inherently more adaptable to API-based LLM, whereas ProFS modifies embeddings, requiring access to internal model representations. These distinctions highlight BART’s suitability for real-world applications where direct modification of model parameters is not feasible.
Given that we are evaluating API-based LLM, we implemented a post-generation correction approach that does not require access to the model’s internal representations. 

Our experiments were conducted on an A100 GPU, ensuring high computational efficiency, and making Corrective-BART a scalable and effective alternative for real-world detoxification while maintaining response quality and usability.

On the other hand, despite its theoretical strength, the ProFS-based method exhibits several limitations that impact its real-world applicability : 
\begin{itemize} 

\item It requires direct access to the LLM's parameters and internal representations, making it less compatible with black-box LLM accessed via API. 

\item The method is computationally expensive, as it relies on PCA to identify toxic subspaces, which demands substantial training data and significant processing time, limiting its scalability for real-time applications.

\item Another key limitation is the risk of semantic distortions, as ProFS operated at the embedding level rather than the text level. Since embeddings encode meaning in high-dimensional space, removing a toxic subspace may inadvertently remove neutral or contextually important information.

\item The ProFs approach struggles with generalization, as it assumes that toxicity can be effectively captured through a fixed set of embedding directions (principal components), while toxic language is highly contextual and evolves dynamically across different topics and linguistic structures.  

\item The method is dependent on predefined toxicity classifiers to label toxic embeddings before training the projection filter, meaning any biases or inaccuracies in these classifiers will propagate into the detoxification process, potentially leading to biased or ineffective filtering. 

\end{itemize}

In contrast, the BART-Corrective Model we propose provides a more practical, efficient, adaptable, and scalable solution for detoxification compared to the ProFS approach that alters embeddings and requires model access. BART reprocesses text responses, preserving context, fluency, and meaning, making it fully compatible with API-based LLM. Unlike ProFS, which relies on dataset similarity, BART dynamically generates alternative outputs, making it more robust to unseen toxicity.  
BART is also faster and computationally lighter, as it eliminates embedding extraction, PCA, and vector projections, enabling real-time operation. 
\section{Conclusion}
\label{Conclusion}
Large Language Models have raised concerns about the safety and security of their generated content, despite their unparalleled text-generation capabilities in various applications. In this paper, a data-centric approach based on a BART-Corrective Model is proposed to improve the safety and security of text generation by significantly lowering the toxicity and security threat levels. The proposed efficient corrective model can be integrated with various LLM and Compound AI systems. In future work, we plan to assess the performance of the proposed correction model with other baseline LLM. We also plan to enrich our approach by addressing toxicity and jail-breaks within a RAG's retrieval system. 


\end{document}